%Paper: hep-th/9506209
%From: "Jose Wudka" <wudka@phyun5.ucr.edu>
%Date: Fri, 30 Jun 1995 16:14:11 -0700

\input epsf
%
%%%
%                     DEFINITION OF SOME SYMBOLS
                              %
%
%%%
%
\def\and{{\it\&}}
\def\half{{1\over2}}

\def\to{\rightarrow}

\def\gesim{\,{\raise-3pt\hbox{$\sim$}}\!\!\!\!\!{\raise2pt\hbox{$>$}}\,}
\def\lesim{\,{\raise-3pt\hbox{$\sim$}}\!\!\!\!\!{\raise2pt\hbox{$<$}}\,}
\def\boldoverdot{\,{\raise6pt\hbox{\bf.}\!\!\!\!\>}}

\def\im{{\bf Im}}
\def\ie{{\it i.e.}}

\def\etal{{\it et. al.}}

\def\ccal{{\cal C}}
\def\dcal{{\cal D}}

\def\jcal{{\cal J}}
\def\kcal{{\cal K}}
\def\lcal{{\cal L}}

\def\ocal{{\cal O}}

\def\tcal{{\cal T}}

\def\pp{{\bf p}}

\def\xx{{\bf x}}
\def\yy{{\bf y}}

\def\BB{{\bf B}}

\def\EE{{\bf E}}

\def\JJ{{\bf J}}

\def\ssb{spontaneous symmetry breaking}
\def\vev{vacuum expectation value}

\def\lhs{left hand side\ }
\def\rhs{right hand side\ }

\def\tr{ \hbox{tr}}

\def\diag{\hbox{\diag}}

%

%
%%%
%                     DEFINITION OF SOME MACROS
                               %
%
%%%
%
\def\doubleundertext#1{
{\undertext{\vphantom{g}#1}}\par\nobreak\vskip-\the\baselineskip\vskip 2pt%
\noindent\ \undertext{\phantom{#1} \hbox to .5in{}}}
\def\noblackbox{\overfullrule=0pt}
\def\leaderfill{\leaders\hbox to 1 em{\hss.\hss}\hfill}
\def\inbox#1{\vbox{\hrule\hbox{\vrule\kern5pt
     \vbox{\kern5pt#1\kern5pt}\kern5pt\vrule}\hrule}}
\def\sqr#1#2{{\vcenter{\hrule height.#2pt
      \hbox{\vrule width.#2pt height#1pt \kern#1pt
         \vrule width.#2pt}
      \hrule height.#2pt}}}
\def\square{\mathchoice\sqr56\sqr56\sqr{2.1}3\sqr{1.5}3}
\def\today{\ifcase\month\or
  January\or February\or March\or April\or May\or June\or
  July\or August\or September\or October\or November\or December\fi
  \space\number\day, \number\year}
\def\pmb#1{\setbox0=\hbox{#1}%
  \kern-.025em\copy0\kern-\wd0
  \kern.05em\copy0\kern-\wd0
  \kern-.025em\raise.0433em\box0 }
\def\up#1{^{\left( #1 \right) }}
\def\lowti#1{_{{\rm #1 }}}
\def\inv#1{{1\over#1}}
\def\su#1{{SU(#1)}}
\def\ui{U(1)}
\def\antes{}
\def\despues{.}
%

% which stands for ``double superscript square''
%
\def\sumprime_#1{\setbox0=\hbox{$\scriptstyle{#1}$}
  \setbox2=\hbox{$\displaystyle{\sum}$}
  \setbox4=\hbox{${}'\mathsurround=0pt$}
  \dimen0=.5\wd0 \advance\dimen0 by-.5\wd2
  \ifdim\dimen0>0pt
  \ifdim\dimen0>\wd4 \kern\wd4 \else\kern\dimen0\fi\fi
\mathop{{\sum}'}_{\kern-\wd4 #1}}
\def\sumbiprime_#1{\setbox0=\hbox{$\scriptstyle{#1}$}
  \setbox2=\hbox{$\displaystyle{\sum}$}
  \setbox4=\hbox{${}'\mathsurround=0pt$}
  \dimen0=.5\wd0 \advance\dimen0 by-.5\wd2
  \ifdim\dimen0>0pt
  \ifdim\dimen0>\wd4 \kern\wd4 \else\kern\dimen0\fi\fi
\mathop{{\sum}''}_{\kern-\wd4 #1}}
\def\sumtriprime_#1{\setbox0=\hbox{$\scriptstyle{#1}$}
  \setbox2=\hbox{$\displaystyle{\sum}$}
  \setbox4=\hbox{${}'\mathsurround=0pt$}
  \dimen0=.5\wd0 \advance\dimen0 by-.5\wd2
  \ifdim\dimen0>0pt
  \ifdim\dimen0>\wd4 \kern\wd4 \else\kern\dimen0\fi\fi
\mathop{{\sum}'''}_{\kern-\wd4 #1}}
%
% these are to give some flexibility in modifying the chapter label
%
\newcount\chapnum
\def\clearchap{\chapnum=0}
\def\chap#1{\clearsect\clearprob
\global\advance\chapnum by 1 \par\vskip .5 in\par%
\centerline{{\bigboldiii\antes\the\chapnum\despues\ #1}}}
\newcount\sectnum
\def\clearsect{\sectnum=0}
\def\sect#1{\clearprob\global\advance\sectnum by 1 \par\vskip .25 in\par%
\noindent{\bigboldii\the\chapnum.\the\sectnum:\ #1}\nobreak}
\def\subsect#1{\par\vskip .125 in \par \noindent{\bigboldi \undertext{#1}}}
\newcount\yesnonum
\def\clearyesno{\yesnonum=0}
\def\verify{\global\advance\yesnonum by 1{\bf (VERIFY!!)}}
\def\tocheck{\par\vskip 1 in{\bigboldv TO VERIFY: \the\yesnonum\ ITEMS.}}
\newcount\notenum

\def\noteeye{\hbox{{$\quad(\!(\subset\!\!\!\!\bullet\!\!\!\!\supset)\!)\quad$}}}

\def\note#1{\global\advance\notenum by 1{\bf \noteeye #1 \noteeye } }
\def\noteout{\par\vskip 1 in{\bigboldiv NOTES: \the\notenum.}}
\newcount\borrownum

\def\borrow{\global\advance\borrownum by 1{\bigboldi BORROWED BY:\ }}
\def\borrowed{\par\vskip 0.5 in{\bigboldii BOOKS OUT:\ \the\borrownum.}}
\newcount\refnum

\def\ref#1{\global\advance\refnum by 1\item{\the\refnum.\ }#1}
\def\stariref#1{\global\advance\refnum by 1\item{%
               {\bigboldiv *}\the\refnum.\ }#1}
\def\stariiref#1{\global\advance\refnum by 1\item{%
               {\bigboldiv **}\the\refnum.\ }#1}
\def\stariiiref#1{\global\advance\refnum by 1\item{
               {\bigboldiv ***}\the\refnum.\ }#1}
\newcount\probnum
\def\clearprob{\probnum=0}
\def\prob{\global\advance\probnum by 1 {\medskip $\triangleright$\
\undertext{{\sl Problem}}\ \the\chapnum.\the\sectnum.\the\probnum.\ }}
\newcount\probchapnum

\def\probchap{\global\advance\probchapnum by 1 {\medskip $\triangleright$\
\undertext{{\sl Problem}}\ \the\chapnum.\the\probchapnum.\ }}
\def\undertext#1{$\underline{\smash{\hbox{#1}}}$}
%

%
% THE FOLLOWING DEFINITION CAN BE USED WITH PHYZZX ONLY!!!!
%
%

\def\UCR{
\address
{{\it Department of Physics\break
                  University of California at Riverside\break
                  Riverside, California 92521--0413; U.S.A. \break
                  \bit}}}

\def\bit{{E{\rm-}Mail address{\rm:} jose.wudka{\rm@}ucr{\rm.}edu}}
%
%
%%%
%                     DEFINITION OF SOME FONTS
                                %
%
%%%
%

 % (this is an old definition)
 % (this is an old definition)

%

%

%
\font\bigboldi=cmbx10 scaled\magstep1
\font\bigboldii=cmbx10 scaled\magstep2
\font\bigboldiii=cmbx10 scaled\magstep3
\font\bigboldiv=cmbx10 scaled\magstep4
\font\bigboldv=cmbx10 scaled\magstep5
\font\small=cmr8
\font\smalli=cmr8 scaled\magstep1
\font\smallii=cmr8 scaled\magstep2

\font\smallv=cmr8 scaled\magstep5
\font\eightrm=cmr8
%
%
%%%
%                      SET UP THE BEGINNING OF A DOCUMENT
                     %
%
%%%
%
\clearchap
\clearyesno
\headline={\ifnum\pageno>0   {\smalli \title (\today)} \hfil {\small Page
\folio } \else\hfil\fi}
%
%
%            TWO COLUMN MACROS
%
%
\newdimen\fullhsize
\newdimen\fullvsize
\newbox\leftcolumn
\def\fullline{\hbox to\fullhsize}
\gdef\twocol{\fullhsize=9.75in
\hsize=4.5in
\vsize=7in
\advance\hoffset by -.5 in
\def\makeheadline{\vbox to 0pt{\vskip-.4in
  \fullline{\vbox to8.5pt{}\the\headline}\vss}
   \nointerlineskip}
\def\makefootline{\baselineskip=24pt
    \fullline{\the\footline}}
\let\lr=L
\output{\if L\lr
   \global\setbox\leftcolumn=\columnbox \global\let\lr=R
  \else \doubleformat \global\let\lr=L\fi
        \ifnum\outputpenalty>-2000 \else\dosupereject\fi
}
\def\doubleformat{\shipout\vbox{\makeheadline
     \fullline{\box\leftcolumn\hfil\columnbox}\makefootline
     }\advancepageno}
\def\columnbox{\leftline{\pagebody}}
\nopagenumbers
\hfuzz=3pt}
\def\twocolphyzzx{
\def\papersize{\hsize=28pc \fullhsize 9.8in \vsize=6in
   \hoffset=-1 in \voffset=0 in
   \advance\hoffset by .5in \advance\voffset by .5 in
   \skip\footins=\bigskipamount \singlespace }
\Tenpoint % This is a new default
\let\lr=L
\output={
    \if L\lr
      \global\setbox\leftcolumn=\columnbox
      \global\let\lr=R
    \else
      \doubleformat
      \global\let\lr=L
    \fi
    \ifnum\outputpenalty>-20000
    \else\dosupereject\fi
}
\def\fullline{\hbox to \fullhsize}
\def\doubleformat{\shipout\fullline{\box\leftcolumn\hfil\columnbox}}
\def\columnbox{\vbox{\leftline{\pagebody}\makefootline}\advancepageno}
}
\gdef\twocollarge{
\def\papersize{\hsize=56pc \fullhsize 19.6in \vsize=12in
   \hoffset=-2 in \voffset=0 in
   \advance\hoffset by 1in \advance\voffset by 1 in
   \skip\footins=\bigskipamount \singlespace }
\def\makefootline{\baselineskip=24pt \fullline{\the\footline}}
\let\lr=L
\output{\if L\lr
   \global\setbox\leftcolumn=\columnbox \global\let\lr=R %\advancepageno
  \else \doubleformat \global\let\lr=L\fi
        \ifnum\outputpenalty>-2000 \else\dosupereject\fi
}
\def\doubleformat{\shipout\vbox{\makeheadline
     \fullline{\box\leftcolumn\hfil\columnbox}\makefootline
     }\advancepageno}
\def\columnbox{\leftline{\pagebody}}
%\nopagenumbers
\def\makefootline{\bigskip\fullline{\the\footline}}
}
%
%
%%%%%%%%%%%%%%%%%%%%%%%%%%%%%%%%%%%%%%%%%%%%%%%%%%%%%%%%%%%%%%%%%%%%%%%%%%%%%%%
%
%       One and two colum defintions for preprints
%
% NOTE: the quantity \colnumber is there in case I need to change things
%       from the one column to the two column formats. For example the
%       layout of some files may need to be changed. This explains the
%       macro \selectfile. Its usage is \selectfile{file1}{file2}
%       inputs file1 in the one column format and file2 in the two
%       column format
%
%%%%%%%%%%%%%%%%%%%%%%%%%%%%%%%%%%%%%%%%%%%%%%%%%%%%%%%%%%%%%%%%%%%%%%%%%%%%%%%
%

%
%

%
%
\def\IIcollarge{
\newlinechar=`\^^J
\immediate\write16{^^J LARGE TWO COLUMN OUTPUT  ^^J}
\immediate\write16{^^J USE dvips -t landscape -x 500 ^^J}
\def\mycolnumber{2}
\input phyzzx
\Mypoint
\twocollarge
\PHYSREV
\paperfootline={\hss\ifp@genum\seventeenrm\folio\hss\fi}
\Mypoint
\def\square{\mathchoice\sqr56\sqr56\sqr{2.1}3\sqr{1.5}3}
\def\vev{vacuum expectation value}
\rightline{UCRHEP-T\ucrnum}
{\titlepage
\vskip -.2 in
\title{ {\bigboldv \thetitle }}
\singlespace
\author{\mycp \theauthor }
\abstract \theabstract
\endpage} \frontpagefalse
}

\def\ucrnum{132}
\def\thetitle{Anomalous commutator corrections to sum rules}
\def\theabstract{In this paper we consider the contributions of
anomalous commutators to various QCD sum rules. Using a combination of
the BJL limit with the operator product expansion the results are
presented in terms of the vacuum condensates of gauge invariant
operators. It is demonstrated that the anomalous contributions are no
negligible and reconcile various apparently contradictory calculations.}
\def\theauthor{ Javier P. Muniain and Jos\'e Wudka \UCR}

%\IIcollarge \hfuzz 51pt \baselineskip 30pt

\newlinechar=`\^^J
\immediate\write16{^^J ONE COLUMN OUTPUT  ^^J}
\def\mycolnumber{1}
\input phyzzx
\Twelvepoint
\PHYSREV
\def\square{\mathchoice\sqr56\sqr56\sqr{2.1}3\sqr{1.5}3}
\def\vev{vacuum expectation value}
\rightline{UCRHEP-T\ucrnum}
{\titlepage
\vskip -.2 in
\title{ {\bigboldiii \thetitle}}
\doublespace
\author{\theauthor}
\abstract
\bigskip
\singlespace
\theabstract
\endpage}  %\PHYSREV
\hfuzz 4 pt \doublespace%\baselineskip 15 pt
\def\whatjournal{P}
\newlinechar=`\^^J

\def\ordernpb#1#2#3{{\bf#1} (#3) #2}
\if P\whatjournal {\global\def\order#1#2#3{\orderprd{#1}{#2}{#3}}}
                    \immediate\write16{^^J PRD references ^^J}\else
                   {\global\def\order#1#2#3{\ordernpb{#1}{#2}{#3}}}
                    \immediate\write16{^^J NPB references ^^J}
\fi

\def\ap#1#2#3{{\it Ann. Phys.\ }\order{#1}{#2}{#3}}

\def\npb#1#2#3{{\it Nucl. Phys. {\bf B}}\order{#1}{#2}{#3}}

\def\plb#1#2#3{{\it Phys. Lett. {\bf B}}\order{#1}{#2}{#3}}

\def\pr#1#2#3{{\it Phys. Rev.\ }\order{#1}{#2}{#3}}
\def\prep#1#2#3{{\it Phys. Rep.\ }\order{#1}{#2}{#3}}
\def\prl#1#2#3{{\it Phys. Rev. Lett.\ }\order{#1}{#2}{#3}}

\def\prd#1#2#3{{\it Phys. Rev. {\bf D}}\order{#1}{#2}{#3}}
\def\ptp#1#2#3{{\it Prog. Theo. Phys.\ }\order{#1}{#2}{#3}}

\REF\schwinger{J. Schwinger \prl{3}{269}{1959}.
T. Goto and I. Imamura, \ptp{14}{196}{1955}.}
\REF\pert{
S. Jo, \npb{259}{616}{1985}; \prd{35}{3179}{1987}.
K. Fujikawa, \plb{171}{424}{1986}, \plb{188}{115}{1987}.
M. Kobayashi \etal, \npb{273}{607}{1986}.
S.R. Gautam and D.A. Dicus, \prd{12}{3310}{1975}.
J.M. Cornwall and R.E. Norton, \pr{177}{2584}{1969}.}
\REF\elec{
R. Jackiw and J. Preparata, \prl{22}{975}{1969}, \prl{22}{1162}{1969}
(errata); \pr{185}{1748}{1969}.
S.L. Adler and Wu-Ki Tung, \prl{22}{978}{1969}; \prd1{2846}{1970}.}
\REF\bjl{J.D. Bjorken, \pr{148}{1467}{1966}.
K. Johnson and F.E. Low \ptp{\hbox{Suppl.} 37-38}{74}{1966}.}
\REF\jackiwgross{For a reference see R. Jackiw in {\sl Lectures on
Current Algebra and Its Applciations}, by S.B. Treiman, R. Jackiw and D.
J. Gross, (Princeton 1972).}
\REF\crewtherold{R.J. Crewther, \prl{28}{1421}{1972}.}
\REF\nonpert{S.L. Adler \etal, \prd{6}{2982}{1972}}
\REF\wilson{K.G. Wilson, \pr{179}{1499}{1969}.}
\REF\svz{M. A. Shifman, A. I. Vainshtein and V. I. Zakharov,
\npb{166}{493}{1980}.}
\REF\jo{S. Jo, Ref. \pert.}
\REF\thooft{G. 't Hooft, \prl{37}{8}{1976}; \prd{14}{3432}{1976}.}
\REF\crewther{
R.J. Crewther, \plb{70}{349}{1977}.
G.A. Christos, \prep{116}{251}{1984}.}
\REF\mottola{E. Mottola, \prd{21}{3401}{1980}.}
\REF\thooftreview{ G. 't Hooft \prep{142}{357}{1986} }
\REF\gl{
J. Gasser and H. Leutwyler, \npb{250}{465}{1985}; \ap{158}{142}{1984}.
J. Gasser \etal, \npb307 (1988) 779.
A. Pich, lectures presented at the {\it V Mexican School of
          Particles and Fields}, Guanajuato, M\'exico, Dec. 1992.}
\def\slacpub{\afterassignment\slacp@b\toks@}
\def\UCRpub{\afterassignment\UCRp@b\toks@}
\def\slacp@b{\Pubnum=\expandafter{UCD--\the\toks@}}
\def\UCRp@b{\Pubnum=\expandafter{UCRHEP-T\the\toks@}}

\def\memohead{\line{\quad\fourteenrm SLAC MEMORANDUM\hfil
       \twelverm\the\date\quad}}
\def\memorule{\par \medskip \hrule height 0.5pt \kern 1.5pt
   \hrule height 0.5pt \medskip}
\def\SLACHEAD{\setbox0=\vtop{\baselineskip=10pt
     \ialign{\eightrm ##\hfil\cr
        \slacbin\cr
        P.~O.~Box 4349\cr
        Stanford, CA 94305\cropen{1\jot}
        (415) 854--3300\cr }}
   \setbox2=\hbox{\caps Stanford Linear Accelerator Center}%
   \vbox to 0pt{\vss\centerline{\seventeenrm STANFORD UNIVERSITY}}
   \vbox{} \medskip
   \line{\hbox to 0.7\hsize{\hss \lower 10pt \box2 \hfill }\hfil
         \hbox to 0.25\hsize{\box0 \hfil }}\medskip }

\FromAddress={\crcr \slacbin \cr
   \P.\ O.\ Box 4349\cr Stanford, California 94305\cr }
\def\slacbin{SLAC\ifx\binno\relax \else , Bin \binno \fi }
\def\binno{81}
\VOFFSET=33pt
\papersize
\catcode`\@=11 % This allows us to modify PLAIN macros.
%%%%%%%%%%%%%%%%%%%%%%%%%%%%%%%%%%%%%%%%%%%%%%%%%%%%%%%%%%%%%%%%
%  Now comes the graphic package.
%  This version is rather primitive
%
\newwrite\figscalewrite
\newif\iffigscaleopen
\newif\ifgrayscale
\newif\ifreadyfile

\def\parsefilename{\ifreadyfile \else
    \iffigscaleopen \else \gl@bal\figscaleopentrue
       \immediate\openout\figscalewrite=\jobname.scalecon \fi
    \toks0={ }\immediate\write\figscalewrite{%
       \the\p@cwd \the\toks0 \the\p@cht \the\toks0 \the\picfilename }%
    \expandafter\p@rse \the\picfilename..\endp@rse \fi }
\def\p@rse#1.#2.#3\endp@rse{%
   \if*#3*\dop@rse #1.1..\else \if.#3\dop@rse #1.1..\else
                                \dop@rse #1.#3\fi \fi
   \expandafter\picfilename\expandafter{\n@xt}}
\def\dop@rse#1.#2..{\count255=#2 \ifnum\count255<1 \count255=1 \fi
   \ifnum\count255<10  \edef\n@xt{#1.PICT00\the\count255}\else
   \ifnum\count255<100 \edef\n@xt{#1.PICT0\the\count255}\else
                       \edef\n@xt{#1.PICT\the\count255}\fi\fi }
\def\redopicturebox{\edef\picturedefinition{\ifgrayscale
     \special{insert(\the\picfilename)}\else
     \special{mergeug(\the\picfilename)}\fi }}
%
%%%%%%%%%%%%%%%%%%%%%%%%%%%%%%%%%%%%%%%%%%%%%%%%%%%%%%
% Few miscellaneous macros
%

%

%%

%%

%
%%%%%%%%%%%%%%%%%%%%%%%%%%%%%%%%%%%%%%%%%%%%%%%%%%%%%%
% Fonts for large two column macro
%
\font\myrmi=cmr10 scaled\magstep5

\font\mybfi=cmbx10 scaled\magstep5

\font\myii=cmmi10 scaled\magstep5     \skewchar\seventeeni='177
      \skewchar\seventeeni='177
\font\mysyi=cmsy10 scaled\magstep5    \skewchar\seventeensy='60
\font\mysyi=cmsy10 scaled\magstep4    \skewchar\seventeensy='60
\font\myexi=cmex10 scaled\magstep5

\font\mysli=cmsl10 scaled\magstep5

\font\myiti=cmti10 scaled\magstep5

\font\mytti=cmtt10 scaled\magstep4
\font\mytt=cmtt10 scaled\magstep3
\font\mycpi=cmcsc10 scaled\magstep5
\font\mycp=cmcsc10 scaled\magstep4
%
%%%%%%%%%%%%%%%%%%%%%%%%%%%%%%%%%%%%%%%%%%%%%%%%%%%%%%%%%%
%
\def\seventeenf@nts{\relax
    \textfont0=\seventeenrm          \scriptfont0=\twelverm
      \scriptscriptfont0=\ninerm
    \textfont1=\seventeeni           \scriptfont1=\twelvei
      \scriptscriptfont1=\ninei
    \textfont2=\seventeensy          \scriptfont2=\twelvesy
      \scriptscriptfont2=\ninesy
    \textfont3=\seventeenex          \scriptfont3=\fourteenex
      \scriptscriptfont3=\twelveex
    \textfont\itfam=\seventeenit     \scriptfont\itfam=\twelveit
    \textfont\slfam=\seventeensl     \scriptfont\slfam=\twelvesl
    \textfont\bffam=\seventeenbf     \scriptfont\bffam=\twelvebf
      \scriptscriptfont\bffam=\ninebf
    \textfont\ttfam=\mytt
    \textfont\cpfam=\mycp }
\def\seventeenpoint{\seventeenf@nts \samef@nt \b@gheight=17pt \setstr@t }
\def\myf@nts{\relax
    \textfont0=\myrmi          \scriptfont0=\seventeenrm
      \scriptscriptfont0=\twelverm
    \textfont1=\myii           \scriptfont1=\seventeeni
      \scriptscriptfont1=\twelvei
    \textfont2=\mysyi          \scriptfont2=\seventeensy
      \scriptscriptfont2=\twelvesy
    \textfont3=\myexi          \scriptfont3=\seventeenex
      \scriptscriptfont3=\twelveex
    \textfont\itfam=\myiti     \scriptfont\itfam=\seventeenit
    \textfont\slfam=\mysli     \scriptfont\slfam=\seventeensl
    \textfont\bffam=\mybfi     \scriptfont\bffam=\seventeenbf
      \scriptscriptfont\bffam=\twelvebf
    \textfont\ttfam=\mytti
    \textfont\cpfam=\mycpi
\chapterfontstyle={\mybfi}
\def\smallii{\smallv}
}
\def\mypoint{\myf@nts \samef@nt \b@gheight=18pt \setstr@t }
\newif\ifmy@
\def\Mypoint{\mypoint\my@true\twelv@false\spaces@t}
\def\Tenpoint{\tenpoint\my@false\twelv@false\spaces@t}
\def\Twelvepoint{\twelvepoint\my@false\twelv@true\spaces@t}
\def\spaces@t{\rel@x
\ifmy@
    \ifsingl@\subspaces@t8:7;\else\subspaces@t9:5;\fi
  \else
\iftwelv@
    \ifsingl@\subspaces@t3:4;\else\subspaces@t1:1;\fi
  \else
    \ifsingl@\subspaces@t3:5;\else\subspaces@t4:5;\fi
\fi \fi
\ifdoubl@ \multiply\baselineskip by 5 \divide\baselineskip by 4 \fi
}
\def\normalbaselines{ \baselineskip=\normalbaselineskip
   \lineskip=\normallineskip \lineskiplimit=\normallineskip
   \iftwelv@ \else
   \ifmy@
\multiply\baselineskip by 5 \divide\baselineskip by 4
     \multiply\lineskiplimit by 5 \divide\lineskiplimit by 4
     \multiply\lineskip by 5 \divide\lineskip by 4
\else
\multiply\baselineskip by 4 \divide\baselineskip by 5
     \multiply\lineskiplimit by 4 \divide\lineskiplimit by 5
     \multiply\lineskip by 4 \divide\lineskip by 5 \fi \fi}
\def\titlestyle#1{\par\begingroup \titleparagraphs
\ifmy@\mypoint\else\iftwelv@\fourteenpoint\else\twelvepoint\fi\fi
\noindent #1\par\endgroup }
\def\refout{\par\penalty-400\vskip\chapterskip
   \spacecheck\referenceminspace
   \ifreferenceopen \Closeout\referencewrite \referenceopenfalse \fi
   \line{\ifmy@\myrmi\else\iftwelv@\fourteenrm\else\twelverm\fi\fi
   \hfil REFERENCES\hfil}\vskip\headskip
   \input \jobname.refs
   }
\def\figout{\par\penalty-400
   \vskip\chapterskip\spacecheck\referenceminspace
   \iffigureopen \Closeout\figurewrite \figureopenfalse \fi
   \line{\ifmy@\myrmi\else\iftwelv@\fourteenrm\else\twelverm\fi\fi
   \hfil FIGURE CAPTIONS \hfil}\vskip\headskip
   \input \jobname.figs
   }
\def\tabout{\par\penalty-400
   \vskip\chapterskip\spacecheck\referenceminspace
   \iftableopen \Closeout\tablewrite \tableopenfalse \fi
   \line{\ifmy@\myrmi\else\iftwelv@\fourteenrm\else\twelverm\fi\fi
   \hfil TABLE CAPTIONS \hfil}\vskip\headskip
   \input \jobname.tabs
   }
\def\address#1{\par\kern 5pt\titlestyle{\it #1}}
  \def\Vfootnote#1{%
      \insert\footins%
      \bgroup%
         \interlinepenalty=\interfootnotelinepenalty%
         \floatingpenalty=20000%
         \singl@true\doubl@false%
         \ifmy@\seventeenpoint\spaces@t
         \smallskip
         \else\Tenpoint\fi%
         \splittopskip=\ht\strutbox%
         \boxmaxdepth=\dp\strutbox%
         \leftskip=\footindent%
         \rightskip=\z@skip%
         \parindent=0.5%
         \footindent%
         \parfillskip=0pt plus 1fil%
         \spaceskip=\z@skip%
         \xspaceskip=\z@skip%
         \footnotespecial%
         \Textindent{#1}%
         \footstrut%
         \futurelet\next\fo@t%
   }
\paperfootline={\hss\ifmy@\seventeenrm\folio\hss%
\else\iffrontpage\else\ifp@genum\tenrm\folio\hss\fi\fi\fi}
\catcode`\@=12 % at signs are no longer letters

\noblackbox

\def\vevof#1{{\left\langle 0 \left| #1 \right| 0 \right\rangle}}

\chapter{Introduction}
The use of canonical commutators in the evaluation of current algebra
relations has produced many results whose effects are directly
measurable. Still in many cases the canonical evaluation of the
commutators is ill defined, as clearly exemplified by Schwinger's
calculation~\refmark\schwinger\ of $ \vevof{ \left[
J^0( 0 , \xx ) , J^i( 0 , \yy ) \right] } $
for a conserved
current $ J^\mu $. If a fermionic current of the type $ \bar \psi
\gamma^\mu \psi $ is replaced in the above expression, and canonical
commutation relations are used the above expression vanishes. In
contrast, using general principles (such as Lorentz covariance and
the absence of negative norm gauge-invariant states), the above \vev\ is
seen to be non-zero.

This fact has been used repeatedly
(although somewhat sporadically) in the calculations of anomalous (\ie,
non-canonical) contributions to various commutators, especially in the
context of anomalous theories~\refmark{\pert}. Similar effects have also
been shown to modify the (canonically obtained) properties of the
electroproduction sum rules~\refmark{\elec}.

Faced with these problems in the canonical evaluation of commutation
relations an alternative definition of the commutators was proposed by
Bjorken and by Johnson and Low~\refmark\bjl. This definition preserves
all the desirable features of the theory, is well defined and coincides
with the canonical results whenever the latter are also well defined.

The starting point of the Bjorken-Johnson-Low (BJL) definition of the
commutator of two operators $A$ and $B$ is the time ordered product $ T
( A B ) $, presented as a function of the momentum transfer $p$.
One then obtains the Laurent expansion of this operator in $p^0$ (the
energy transfer). The term proportional to $ 1/p^0$ is identified as
(the Fourier transform of) the equal time commutator~\foot{It is of course
possible for this quantity to be divergent.} Terms in this expansion
containing positive powers of $p^0$ are associated with the
covariantizing of the time-ordered product~\refmark\jackiwgross\ and can
be ignored.

The applications of this method have been largely restricted to
perturbation theory (see however Refs. \crewtherold,\nonpert).
On the other hand, many interesting
applications of current algebra reside in the area where perturbation
theory cannot be applied. In order to use the BJL definition in a
wider range of situations we first note that the commutator is obtained
by studying the relevant time-ordered products in the limit of large
energy transfers, and, therefore, that an operator product
expansion~\refmark\wilson\ (OPE) is appropriate. The procedure which we
follow is therefore to perform an OPE of the said
time-ordered product, to then use renormalization group arguments to
determine the high-energy behavior of the coefficient functions and
thus extract the terms that contribute to the commutator. The result is
then expressed in terms of the residues of the coefficient
functions multiplied by the matrix elements of the local operators
appearing in the OPE\foot{This is reminiscent of the results obtained
using sum rules~\refmark\svz.}. In these calculations all
symmetries of the theory are manifest, and so the resulting
commutator will also respect them. A similar method was proposed
by Crewther many years
ago~\refmark\crewtherold\ but was not developed significantly.

This paper is organized as follows. In the following section we describe
the method in detail. Section 3 presents a comparison of the present
method with some explicit perturbative calculations in 1+1 dimensions.
In sections 4 and 5
we consider the anomalous commutator modifications to the current algebra
approach to the $ \ui_A $ problem. Section 6 presents some explicit
calculations pertaining the general arguments presented in sections 4
and 5. Parting comments are presented in
section 7. The appendix contains the comparison of the present method
with the results of perturbation theory for a 3+1 dimensional model.

\chapter{Description of the method}

In this section we develop a useful technique which allows us to extract
information about non-canonical contributions to equal time commutators
without going through the lengthy steps of
loop calculations involving triangle, box and even pentagon
diagrams. This technique, which does not rely on perturbation theory, is
based on the Bjorken-Johnson-Low definition of the equal time
commutators and on the operator product expansion (OPE).

According to the BJL limit prescription, the
definition of the commutator of two operators is obtained from the high
energy behavior of Green's functions
$$ \eqalign{
 \lim_{p_0 \to \infty} & i p_0 \int d^n x \;e^{ - i p \cdot x}
\left\langle
\alpha \left| T A ( x/2 ) B ( - x/2 ) \right| \beta \right\rangle = \cr &
\qquad =
\int d^{n-1} x\; e^{ i \pp \cdot \xx } \left\langle
\alpha \left| \left[ A ( 0 , \xx/2 ) \; , \; B ( 0 , - \xx/2 )
\right] \right| \beta
\right\rangle , \cr} \eqn\bjllim $$
where $p_0$ is the time-like component of the four momentum~\foot{Double
commutators can be defined using a straightforward generalization
involving a double limit.}. The time ordered
product $T$ is not Lorentz invariant and differs from the corresponding
covariant Green's function by terms involving delta functions of $x_0$
and its derivatives (corresponding to a polynomial in $p_0$
in momentum space). If the \lhs\ of \bjllim\ is evaluated using
covariant perturbation theory then all polynomials in $ p_0 $ should be
dropped. The covariant time-ordered product will be denoted by $T^*$

Since we are interested in the large momentum behaviour of $ \langle
\alpha | T^* A B | \beta \rangle $, it is appropriate to express this
object as a sum of local operators (OPE) where the coefficient functions
summarize the $ p_0 \rightarrow \infty $ behaviour,
$$ \int d^n x \;e^{- i p \cdot x} \left\langle
\alpha \left| T^* A ( x/2 ) B ( - x/2 ) \right| \beta \right\rangle =
\sum_r c_r(p)\; \langle \alpha | \ocal_r | \beta \rangle ,\eqn\ope $$
where the local operators $ \ocal_r $ are evaluated at $ x =0 $.
Taking the BJL limit of the previous expression we find
$$ \eqalign{
\int d^{n-1} x\; e^{ i \pp \cdot \xx } & \left\langle
\alpha \left|
\left[ A ( 0 , \xx/2 ) \; , \; B ( 0 , - \xx/2 ) \right] \right| \beta
\right\rangle \cr & \qquad
= \sum_r \lim_{ p_0 \rightarrow \infty } \left[ i p_0
c_r(p) \right] \; \langle \alpha | \ocal_r | \beta \rangle , \cr }
\eqn\opebjl $$
where, as mentioned above, all terms in the $ c_r $ growing as a power
of $ p_0 $ can be dropped.

It is well known that the matching of dimensions of the operators
$ T ( A B ) $ and $ \ocal_r$ in the OPE must take into account the anomalous
dimensions of these objects. This can be avoided provided the operators
considered are renormalization group invariant, such as the trace
of the energy momentum tensor or the fermion mass terms. Note that even
if $A$ and $B$ are renormalization group invariant, the time ordered
product $ T ( A B ) $ need not have this property. In the most favorable
cases the operators are renormalization group invariant and the
canonical evaluation of the dimensions remains valid.

Another characteristic of the method is that the results are evaluated
in terms of a set of unknown constants, the residues of the
the coefficient functions $ c_r $. For the applications which we
consider this will not be a disadvantage: these constants multiply the
matrix elements $ \langle \alpha | \ocal_r | \beta \rangle $ which, in
most cases cannot be evaluated to all orders in perturbation theory.
Thus the final result will be given in terms of these ``condensates''
multiplied by the said constants.

\chapter{Simple example}

As an application the previous remarks we consider a model containing
fermions coupled to external non-Abelian gauge fields. We then choose $
A = \jcal_\mu^a , \ B = \jcal_\nu^b $, where $ \jcal $ denotes the
right or left-handed, gauge invariant fermionic current, and $a,b$ denote
color indices. Thus we consider $$ \tcal_{ \mu \nu }^{ a b } = \int
d^n x \; e^{ - i p \cdot x } \;  T^* \jcal_\mu^a ( x/2 ) \jcal_\nu^b ( - x/2 )
{}.
 \eqn\defoftcal $$ Similarly we define $$ \ccal_{ \mu \nu } ^{ a b } =
\lim_{ p^0 \rightarrow \infty} i p_0 \tcal_{ \mu \nu }^{ a b }
\eqn\defofccal $$
(where the terms growing like a polynomial in $ p_0$ are dropped as
discussed in the previous section).
In writing the operator expansion of this object we have to pick terms
that have the same dimensionality and that posses same symmetries; in
particular we can restrict ourselves to  gauge invariant operators.

In $n$ dimensions $\tcal \sim \left( \hbox{mass } \right)^{ n-2 }$,
hence we can restrict
ourselves to operators $\ocal_r$ of dimension equal or greater than
$(n-2)$ on the \rhs of the OPE. We will consider here the
1+1 dimensional case, leaving the 1+3 case to the appendix, as it does not
bring up any new ideas or physics, and it is somewhat more involved.

In 1+1 dimensions $ \tcal $ has canonical dimension zero.
Moreover, the Dirac matrices satisfy $ \gamma_5 \gamma^\mu = \epsilon^{
\mu \nu } \gamma_\nu $ which implies that we need only consider the
vector currents, which we denote by $ J_\mu^a $.
The current is given explicitly by
\def\thecurrenti{
J^a_\mu (x,\epsilon) = { i \over 2} \bar\psi \left( x +
{\epsilon \over 2} \right) \gamma_\mu T^a \left[ {\cal P} \exp
\left( \int_{x-\epsilon /2}^{x+\epsilon /2} A^\sigma (y)\,dy_\sigma
\right) \right] \psi \left(x-{\epsilon \over 2}\right)  + \hbox{h.c.} }
$$ \thecurrenti, \eqn\defofj $$
where $T^a$ denote the (anti-hermitian) group generators,~\foot{The
conventions we use are $ \tr (T^a T^b) = - \delta_{ab} $,
$ [ T^a, T^b ] = f_{abc} \, T^c $, $ \tr T^a \{ T^b , T^c \} =
- d_{ a b c } $.} ``h.c'' denotes the Hermitian
conjugate,
and $A^\mu = A_a^\mu\, T^a$ denotes the gauge field (the coupling
constant is absorbed in the definition of $A$).

The OPE of  $\tcal $ (defined now for the vector currents by replacing
$ \jcal \rightarrow J $ in \defoftcal)
is then given by $$ \tcal_{ \mu \nu }^{ a b } =
c_0{}_{ \mu \nu } ^{ a b } {\bf1} +
c_1{}_{ \mu \nu \rho } ^{ a b c } ( p ) J^c{}^\rho +
c_2{}_{ \mu \nu \rho } ^{ a b } ( p ) J^\rho +
\cdots , \eqn\eq $$
where {\bf1} denotes the unit
operator, $ J_\rho^c $ is defined in \defofj, $ J_\rho $ denotes the
singlet vector current $ \sim \bar \psi \gamma_\rho \psi $,
the ellipsis denote terms which will
not contribute once the BJL limit is taken\foot{ Note that the
coefficient functions for the operators
$ \bar \psi T^c \psi$, $ \bar \psi \psi $,
$ \bar \psi T^c \gamma_5 \psi$ and $ \bar \psi \gamma_5 \psi $
will be of the form
$ \sim ( \hbox{mass \ parameter})  / p^2 $ and so will not contribute in
the BJL limit. }; the c-number functions $ c_r
$ must have dimension $ -1 $ for $ r = 1 ,2 $ and $0$ for $ r = 0 $.
Using the fact that $ \tcal $ must be symmetric under $ a
\leftrightarrow b , \ \mu \leftrightarrow \nu , \ p \leftrightarrow -p
$, the most general form of the coefficient functions is~\foot{The
constants $ \xi , \ldots, w$ can be evaluated perturbatively. We will
not need their explicit expressions.} $$ \eqalign{
c_0{}_{ \mu \nu } ^{ a b } = & \delta_{ a b } \left[ \xi g_{ \mu \nu
} + \eta p_\mu p_\nu / p^2 \right] , \cr
c_1{}_{ \mu \nu \rho } ^{ a b c } ( p ) = &
\inv{p^2} d_{a b c } \left[
\alpha \left( g_{ \mu \rho } p_\nu - g_{ \nu \rho } p_\mu \right) +
\beta \left( \epsilon_{\mu \rho } p_\nu - \epsilon_{\nu \rho }
p_\mu \right)
+ \gamma \epsilon_{\mu \nu } p_\rho \right] \cr
& + \inv{p^2} f_{a b c } \left[
{a \over 2} \left( g_{ \mu \rho } p_\nu + g_{ \nu \rho } p_\mu \right) +
{b \over 2} \left( \epsilon_{\mu \rho } p_\nu + \epsilon_{\nu \rho } p_\mu
\right) \right. \cr & \left. \qquad \qquad \qquad
+ c g_{\mu \nu } p_\rho + d { p_\mu p_\nu p_\rho \over p^2 } \right] , \cr
c_2{}_{ \mu \nu \rho } ^{ a b } ( p ) = & \inv{ p^2 } \delta_{ a b } \left[
u \left( g_{ \mu \rho } p_\nu - g_{ \nu \rho } p_\mu \right) +
v \left( \epsilon_{\mu \rho } p_\nu - \epsilon_{\nu \rho } p_\mu \right)
+ w \epsilon_{\mu \nu } p_\rho \right] . \cr } \eqn\eq $$
Denoting by $P$ the spatial component of the momentum the commutator
for the vector currents, obtained by replacing $ \jcal \rightarrow J $
in \defoftcal\ and \defofccal, is
given by~\foot{In our conventions $ \epsilon^{ 0 1 } = + 1 $.}
$$ \eqalign{
\inv i \ccal_{ 0 0 }^{ a b } = & f_{ a b c } \left[ ( a + c + d ) J_0^c + b
J_1^c \right] , \cr
\inv i \ccal_{ 0 1 }^{ a b } = & - \eta \delta_{ a b } P - d_{ a b c } \left[ (
\beta + \gamma ) J_0^c + \alpha J_1^c \right] + \half f_{ a b c } \left(
b J_0^c + a J_1^c \right) \cr & \qquad
-  \delta_{ a b } \left[ (v + w ) J_0 + u J_1 \right] , \cr
\inv i \ccal_{ 1 1 }^{ a b } = & - c f_{ a b c } J_0^c . \cr } \eqn\eq $$
In the case where the theory has only right-handed couplings these
relations imply~\foot{Note that in 1+1
dimensions $ J_R^a{}_{\mu=0} = J_R^a{}_{\mu=1} = J_R^a $.} $ d = 0 $ and
$$ i \left[ J_R^a ( 0 , X/2 ) , J_R^b ( 0 , -X/2 ) \right]
= { i \eta \over 2 } \delta_{ a b } \delta' ( X ) - { a + b \over 2 }
f_{ a b c } J_R^c \delta( X ) \eqn\jrcomm $$
This expression should be compared to those obtained in
Ref. \jo\ which has
the same form, except that $ J_R $ is replaced by $ A_R = A_0 + A_1 $.
The discrepancy can be understood by following the procedure used in Ref.
\jo. What was done was to evaluate various matrix elements of
the commutators and then to exhibit some local operators which have the
same matrix elements. These operators are not unique, however. For
example the matrix elements of $ A_R $ and $J_R $ between the vacuum and
the one gauge-boson state are proportional to each other in the zero
momentum limit (the limit in the case of $ J_R$ is taken
symmetrically, first averaging over the direction of the momentum and
then letting the magnitude go to zero). It is easy to see that the
results of the diagrammatic calculations are consistent with those
presented in Ref.\jo\ when $ A_R $ is replaced by $ - 2 \pi J_R $. It is in
this sense that the above calculation in consistent with the explicit
diagrammatic evaluation (up to the undetermined constants $ \eta $ and $
a+b $ which we do not evaluate at this point.) We also point out that
the above expressions have the expected form when taking the matrix
elements of the commutators for states containing fermions.

The above expressions of the anomalous commutators have the
additional advantage of
being manifestly gauge covariant. The terms proportional to $
\delta ( X ) $ are generated by the matrix elements of the canonical
contribution to the commutator. The only irreducible non-canonical
contribution
is the Schwinger term $ \propto \delta_{ a b } \delta' ( X ) $. We
shall see in the appendix that similar results hold in 3+1 dimensions.

The above results can also be used for calculating the Schwinger term
and seagull for the commutators under consideration. Writing the
expressions using a time-like unit vector $n$~\refmark{\jackiwgross} we
obtain for the commutator of two vector currents, $$ \eqalign{
& i \left[ J_R^a ( x/2 ) , J_R^b ( -x/2 ) \right] \delta( x \cdot n ) =
C^{a b }_{ \mu \nu } \delta\up2(x) + S_{\mu \nu}^{ a b ; \; \alpha }
\partial_\alpha \delta\up2( x ) \cr \hbox{ where \quad }& C^{a b }_{ \mu \nu }
= - f_{ a b c } \left( \left| g_{ \mu \nu } \right| n \cdot J^c + \left|
\epsilon_{ \mu \nu } \right| n_\rho \epsilon^{ \rho \sigma } J^c_\sigma
\right) ; \cr \phantom{ \hbox{ where \quad } } &
S_{\mu \nu}^{ a b ; \; \alpha } = \delta^{ a b } \eta \left|
\epsilon_{ \mu \nu } \right| n_\rho \epsilon^{ \rho \alpha } \cr }
\eqn\eq $$ From this it follows that the corresponding seagull vanishes.

Thus the method is seen to work to lowest order in perturbation theory.
The disadvantage is that the final result is expressed in terms of a few
unknown constants which, if required, can only be obtained doing
detailed calculations. Moreover, for higher orders in perturbative
calculations the anomalous dimensions of the various operators must be
taken into account. We have seen that the apparently gauge
variant results obtained in the literature can be re-interpreted as
generated by the canonical terms in the commutator.

\chapter{Current algebra and the U(1) problem.}

In this section we will consider the effects of anomalous commutators in
the study of the $ \ui_A $ problem. In this area the results obtained
using instanton
calculations~\refmark\thooft\ were criticized~\refmark\crewther\
on the basis of certain
inconsistencies which arise when the commutators involved are evaluated
using canonical expressions. We will see that the relations derived in
Ref. \crewther\ are in general modified due to the anomalous terms in
the commutators; this point is also made in
Refs.~\mottola~and~\thooftreview\ where it is noted that configurations
carrying topological charge affect the pion decay constant
Reference \crewther\ also points out several apparent
contradictions concerning the periodicity of the $ \theta $ angle within
the instanton and the canonical approaches. This problem was
investigated in Refs.~\mottola~and~\thooftreview\ and found to be rooted
in a mis-application of the index theorem for which there are subtleties
connected with \ssb.

These modifications are sufficient to explain the
differences between the two approaches.

To specify the notation we denote by $ J_5^\mu $ the gauge-invariant
anomalous current which, in the presence of $ \ell $ massless flavors,
satisfies $$ \partial_\mu J_5^\mu  = \ell \; \nu ; \quad
\nu = { g^2 \over 16 \pi^2 } F \cdot \tilde F .\eqn\anomeq $$
The charge associated with this current is denoted
by $ Q_5 = \int d^3 x \; J_5^0 $.

In the effective lagrangian description the effects of an instanton
(respectively, an anti-instanton) localized at $x$ is described by a
potential~\refmark{\thooftreview} $
U\lowti{a} $ (respectively, $ U\lowti{a}^* $) with the identification
($ \ell $ is the number of light quark flavors)
$$ \ell \; \nu(x) \leftrightarrow 8 \, \im \; U\lowti{a} ( x )  ,\eqn\iden $$
where $ \nu $ is defined in \anomeq.
The potential $ U\lowti a $ is proportional to a quark-determinant
operator involving all light flavors~\refmark{\thooft,\thooftreview}.

The
problem arises because $ U\lowti{a} $ has chirality $ 2 \ell $
and so the \rhs of \iden\ has non-trivial commutator with the axial
charge (as constructed in the effective theory). In contrast, $ \nu
$ apparently commutes with $ Q_5 $, thus raising questions about the
above identification.
This contradiction can be solved by using the BJL definition of the
commutator between $ \nu $ and $ J^\mu_5 $.

As a fist step we
consider, for example, the vacuum correlator $$
\int d^4 x \;e^{- i p \cdot
x} \vevof{  T^* \nu ( x/2 )  J_5^\mu ( - x/2 ) } = \sum_r c^\mu_r(p)\;
\vevof{ \ocal_r } . \eqn\anomope $$ The lowest dimensional (non-trivial)
operator that contributes to the \rhs\ is the
trace of the energy-momentum tensor which we denote by
 $ \Theta $. We expect the commutator to be a renormalization group
invariant quantity; in this case
the coefficient function associated with $ \Theta $ will have the
form $ c_\Theta^\mu ( p ) = \bar c_\Theta p^\mu/ p^2 $. We hasten to
point out that the OPE is valid only for large $p$ and so one cannot
interpret this form of  $ c_\Theta^\mu ( p ) $ as corresponding to a
massless pseudoscalar excitation. We then obtain
$$ \int d^4 x e^{ i \pp \cdot \xx }
\vevof{ \left[ \nu ( 0 , \xx/2 ) \; ,\;  J_5^0 ( 0 ,
- \xx/2 ) \right] } = i \; \bar c_\Theta \; \vevof
\Theta . \eqn\nujcomi $$ Since we expect both $ \bar c_\Theta $ and $
\vevof \Theta $ to be non-vanishing, it follows that the
commutator of $ \nu $ with $ Q_5 $ is non-trivial also within the
context of QCD~\foot{A straightforward perturbative calculation shows that,
at least to one loop, $ \bar c_\Theta \not= 0 $; see below.}.

When quark masses are included the above equation is modified since more
operators become available. Specifically, one can include on the \rhs\ of
\anomope\ a term containing the operator $ D $, where $$ D = 2 \sum_f
m_f \bar q_f q_f , \eqn\defofD $$ ($m_f$ and $ q_f $ denote the mass and field
associated with the quark of flavor $f$). In this case the corresponding
coefficient function in the OPE \anomope\ tales the form $ c_D^\mu =
\bar c_D p^\mu / p^2 $, and \nujcomi\ becomes
$$ \int d^4 x e^{ i \pp \cdot \xx }
\vevof{ \left[ \nu ( 0 , \xx/2 ) \; ,\;  J_5^0 ( 0 ,
- \xx/2 ) \right] } = i \; \bar c_\Theta \; \vevof
\Theta  + i \bar c_D \; \vevof D . \eqn\nujcomii $$

As we will see below, $ \bar c_{ \Theta , D } $ do not in general
vanish. Hence the commutator between $ \nu $ and $ Q_5 $ receives
non-canonical contributions. Model calculations~\refmark{\thooftreview}
also show that the expression for the said commutator acquires a
non-canonical piece proportional to $ \vevof \Theta $.
The \rhs\ of \nujcomii\ should vanish for massless quarks
in the $ \pp \rightarrow 0 $ limit; this is verified within a specific
model in section 6.

\chapter{Anomalous Ward identities}

In the previous section we remarked that the operator $ \nu $ can
have non-zero commutator with the gauge invariant axial current $ J_5^\mu $;
in particular $ \left[ \nu , Q_5 \right] \not= 0 $. These results are
supported by a straightforward application of the effective lagrangian
proposed in Ref.~\thooftreview.
It is of course possible for the
constant $ \bar c_\Theta $ (and $ \bar c_D $ is $ m_f \not= 0 $)
to vanish, but this would not be consistent with the effective
lagrangian approach. We also point out that $ \nu $ will
mix under renormalization with operators which have non-zero
chirality.

Should the above commutator be different form zero, the anomalous
Ward identities will be modified. Consider then a gauge invariant
operator $ \ocal $ and define $$ \Pi\up\ocal_\mu ( p )  =
\int d^4x \; e^{ - i p \cdot x } \vevof{ \; T \; J_5{}_\mu(x)
\; \ocal(0) } . \eqn\eq $$ The requirement that there be no light isosinglet
pseudoscalars~\refmark{\crewther,\svz}
implies that $ p \cdot \Pi \up\ocal $ will vanish
as $ p \rightarrow 0 $. It follows that, by the definition of the
$T$ symbol, $$ \eqalign{ 0
=& \int d^4 x \; \vevof{ \; T \; \partial \cdot J_5(x) \; \ocal(0) } + \vevof{
\left[ Q_5 , \ocal(0) \right] } \cr
=& \int d^4 x \; \vevof{ \; T \; \Delta(x) \; \ocal(0) } + \int d^4 x \;
\vevof{ \; T \; \ell \nu(x) \; \ocal(0) } + \vevof{ \left[ Q_5 \; , \; \ocal(0)
\right] } \cr
} \eqn\generalidentity $$ where $$ \Delta = 2 i \sum_{ f = 1 }^\ell m_f \bar
q_f \,
\gamma_5 \, q_f \eqn\eq $$
and $ \nu $ is defined in \anomeq; we have assumed that the anomaly
equation, $ \partial \cdot J_5 = \Delta + \ell \nu $, is an operator identity.

Now, following Refs.~\crewther~and~\svz, we consider \generalidentity\ for the
cases $ \ocal = \ell \nu $ and $ \ocal = \Delta $; cancelling the
correlator of $ \Delta $ and $ \nu $ which appears in both these
expressions we obtain $$ \ell^2 \int d^4 x \; \vevof{ \; T \; \nu(x) \; \nu(0)
}
=
\vevof{ \left[ Q_5 \; , \; \Delta(0) - \ell \nu(0) \right] }
+ \int d^4 x \vevof{ \; T \; \Delta(x) \; \Delta(0) }
. \eqn\eq $$

The $T$-product on the \rhs equals the corresponding $T^*$ product.
This is because the approach described in section 2 shows that there are
no Schwinger terms in the equal-time commutator of $ \Delta (x) $ and
$ \Delta(y) $; the corresponding
seagull is therefore zero~\refmark\jackiwgross. The commutator
$ [ Q_5, , \Delta ] $ is proportional to
$ D = 2 \sum m \bar q q $;
we will write
$$ i \left[ Q_5 \; , \; \Delta \right] = 2 ( 1 + \delta ) \; D ,
\qquad D = 2 \sum_{ f = 1 }^\ell m_f \bar q_f q_f\eqn\eq $$
where $ \delta = 0 $ if the commutator is evaluated canonically.
Finally we have
$ i [ Q_5 , \nu ] = \bar c_\Theta \Theta + \bar c_D D $, where
$ \Theta $ denotes
the trace of the energy-momentum tensor.
Thus we obtain, for the case of
two light flavors ($\ell = 2 $),
$$  2 i \int d^4 x \; \vevof{ \; T \; \nu(x) \; \nu(0) }  =
\left( \delta - \bar c_D \right) \vevof D - \bar c_\Theta \vevof\Theta
- 4 f_\pi m_\pi^2 { m_u m_d \over ( m_u + m_d )^2 } . \eqn\eq $$
The fist two terms come from the anomalous commutators,
while the last term is generated by the canonical commutator
and the $ T \Delta \Delta $ contributions as evaluated
in Ref.~\svz.

The above calculation show that in general we can expect deviations
form the canonical expression for the dependence of measurable quantities
on the CP-violating parameter $ \theta $. It is of course possible for
the non-canonical terms to vanish, still explicit perturbative calculations
and effective-lagrangian arguments favor $ \bar c_{\Theta , D } , \;
\delta \not= 0 $.
Since the dependence on $ \theta $ disappears from all physical
observables when one fermion is massless, we expect $ \bar c_\Theta
\vevof\Theta - \delta \vevof D $ to vanish when  any quark  mass
is zero. We can then write
$$ i \int d^4 x \; \vevof{ \; T \; \nu(x) \; \nu(0) }  =
- 2 f_\pi^2 m_\pi^2 ( 1 - \lambda ) { m_u m_d \over ( m_u + m_d )^2 } ,
\eqn\defoflam  $$ for some constant $ \lambda $.
The conditions under which $ \lambda = 1 $ (or even if this is
at all possible) cannot be determined using general arguments.
We will see below that low-energy models of the strong interactions
predict $ 1 - \lambda = O ( 1 ) $ (see below)
so that the estimates of physical
quantities on the trong CP angle $ \theta $ are altered only by a factor
of $O(1)$ (except, of course, in the case $ \lambda = 1 $).

In the following section we present several calculations where the
various coefficients and vacuum condensates are evaluated within explicit
models.

\section{Explicit calculations}

In this section we present several computations. We have evaluated the
OPE coefficients $ \bar c_\Theta , \  \bar c_D $ and $ \delta $ to
lowest order in QCD; we also evaluate  the condensates $ \vevof \Theta $
and $ \vevof D $ in a chiral model of the strong interactions.

\subsect{Perturbative calculations}

We first consider the calculation of $ c_\theta^\mu $.
The OPE of the product $ T^* J^\mu_5 \nu $ contains, to lowest order,
three operators: $D$ defined in \defofD, $ \Theta_{ \mu \nu } $, the
energy-momentum tensor, and $ \Theta $, its trace. For the calculation
at hand we evaluate the matrix element between the vacuum and a two
gluon state, the relevant graphs are presented in figure 1.

\setbox2=\vbox to 2.6 truein{\epsfysize=6 truein\epsfbox[100 -100 712
692]{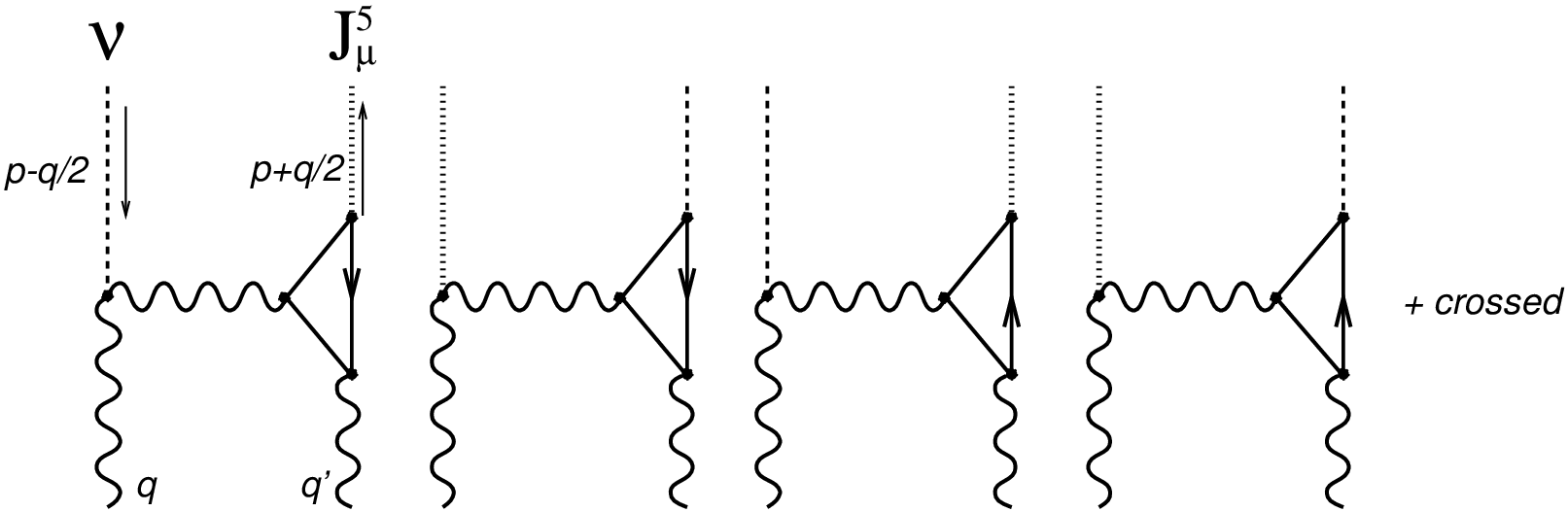}}
\centerline{\box2}

\catcode`\@=11 % This allows us to modify PLAIN macros.

\centerline{\vbox to .05 in{\hsize 4 in \vfill
           {\ifmy@\seventeenpoint\spaces@t\else\Tenpoint\fi
           \textindent{Fig. 1} \global \advance \baselineskip by -12 pt
            \noindent Lowest order diagrams contributing to the OPE
                      coefficient of $ \Theta $ in the operator
                      $ T^* J_\mu^5 \nu $.
           }}}\global \advance \baselineskip by 12 pt

\bigskip
\catcode`\@=12 % at signs are no longer letters

In order to obtain the OPE coefficients in the $
p \rightarrow \infty $ limit we need only consider the diagrams
contracted with the momentum carried by the axial current, \ie, we
multiply each graph by $ ( p + q/2 )^\mu $ (the momentum of the axial
current) and contract the index $ \mu
$. A simple calculation using $$ \Theta = { \beta( \alpha )
\over 4 \alpha } \left( F_{ \mu \nu }^a \right)^2 + \half ( 1 +
\gamma_m ) D ,\eqn\defofTheta$$ where $ \beta $ denotes the beta function
for $\alpha = g^2 / 4 \pi $ and $ \gamma_m $ the mass anomalous
dimension. Assuming the presence of $ \ell $ fermions we obtain
$$ \bar c_\Theta = \ell { \alpha^3 \over \pi^2 \; \beta ( \alpha )
} ; \qquad \beta( \alpha ) = - { 11 - 2 \ell/3 \over 2 \pi } \alpha^2 +
O ( \alpha^3 ) . \eqn\eq $$

We now consider $ \bar c_D $ which can be extracted from the matrix
element of $ T \nu J_5^\mu $ between the vacuum and a two fermion state.
It is easy to
see, however, that the graphs for this matrix element are all $ O (
\alpha^2 ) $; in contrast $ \bar c_\Theta = O ( \alpha ) $. The OPE for
$ T \nu J_5^\mu $ contains both
$ \Theta $, defined in \defofTheta, and $D$. It follows that the
$ O ( \alpha ) $
term in $ \bar c_\Theta $ should be cancelled by a similar term in
$ \bar c_D $, hence $$ \bar c_D = \half \bar c_\Theta \eqn\eq $$

Finally we calculate $ \delta $ by evaluating the matrix element of $
T^* \Delta J_\mu^5 $ between the vacuum and a two fermion state. We skip
for brevity the description of the graphs; the final result is
$$ \delta = - {9 \over 4 } { \alpha \over \pi } \eqn\eq $$ to first
order in $ \alpha $.

these calculations show that, at lest to first order in perturbation
theory, all the anomalous coefficients are finite and non-vanishing as
claimed previously.

\section{Model Calculations}

In order to obtain estimates of the condensates $ \vevof \Theta
$ and $ \vevof D $ we consider a
chiral model which obeys the same symmetries as QCD. In
order to generate Green's functions involving $ \nu $ we modify the QCD
lagrangian by adding a term $ \theta \nu /2 $, with $ \theta $ an
external source (for details see Ref. \gl). The Lagrangian
takes the form~\refmark{\gl} $$ \lcal = - V_0 + V_1 \tr
\partial_\mu U^\dagger \partial^\mu U  + \left( V_2 \tr M U + \hbox{H.c}
\right) + V_3 \partial \phi_0 \cdot \partial \theta + V_4 \left(
\partial \theta \right)^2 \eqn\eq $$ where $V_i = V_i ( \phi_0 + \theta
) $, $ V_{ i \not=2 } $ real, and $ V_i ( \alpha ) = V_i^* ( - \alpha
) $. The meson field, denoted by $ U $, belongs to the unitary field
$ U ( \ell ) $; we will write $ U = \exp\left(i \phi_0 / \ell \right) \; \Sigma
$ with $ \Sigma \in \su\ell $. For $ \ell = 3 $ $ \Sigma $ describer the
pseudoscalar meson octet and $ \phi_0 $ describes the pseudoscalar
isosinglet (the $ \eta' $).

The field $ \Sigma $ describes the usual pseudoscalar meson multiplet (under
$ \su\ell $ flavor); $ \phi_0 $ describes the pseudoscalar singlet (\ie,
the $ \eta $ for $ \ell =2 $, and the $ \eta' $ for $ \ell =3 $).

This model is an accurate representation of QCD at low momentum
transfers, so we will not use it in obtaining the BJL definition
of the commutators (which
involve the $ p^0 \rightarrow \infty $ limit). We can, however, use this
model to evaluate the condensates $ \vevof \Theta $
and $ \vevof D $ and the low momentum limit of $ T^* \nu \nu $.
We will, for simplicity, work in
the $ \su \ell $ symmetric limit where $ M = m {\bf1 } $.

To lowest order in a momentum expansion the correlator $ \vevof T^* \nu
\nu $ can be obtained by replacing $ U $ in  $ \lcal $ by the solution to
the classical equations of motion~\refmark{\gl}. A simple calculation
shows that the lagrangian then takes the form $$ \bar\lcal = \half \theta
\kcal \theta + \hbox{const.} ; \qquad \kcal = - 2 \left[ V_4( 0 ) + \
inv \ell V_1( 0 ) -
V_3( 0 ) \right] \partial_\mu \partial^\mu - { 2 m \over \ell } V_2( 0 )
\eqn\eq $$ which
shows that the vacuum correlator $ \vevof{T^* \nu \nu }
$ is proportional to $m$ in the limit of zero momentum transfer.

The various condensates can also be evaluated within this model.
 From Ref. \gl\ we get $$
\vevof D = - 2 \ell m_\pi^2 f_\pi^2 ; \qquad \vevof \Theta = 4 V_0(0) -
4 \ell m_\pi^2 f_\pi^2 , \eqn\eq $$ where $ m_\pi  $ denotes the
(degenerate) meson mass and $ f_\pi $ the corresponding decay
constant. Perturbative calculations
suggest that $ \bar c_\Theta $ remains non-zero as $ m \rightarrow 0 $,
hence consistency of the OPE with the above expressions for $ T^* \nu
\nu $, $\vevof D $ and $\vevof\Theta $ requires $ V_0(0) $ to vanish as $
m \rightarrow 0 $. It follows that both condensates $\vevof D, \
\vevof\Theta $ vanish in this limit. This result justifies the claims
made at the end of the previous section concerning the behaviour of the
condensates in the zero mass limit. Similar results are obtained using
the (closely related) model of Ref.~\thooftreview

Within this model $ i \int d^4 x \vevof{T^* \nu \nu } = - \half ( m_\pi
f_\pi )^2 $ which corresponds to $ \lambda = 0 $ in \defoflam. Thus the
possibility of having $ \lambda = 1 $ and a dynamical cancellation of
the dependence on the strong CP angle is not realized, at least within
this model.

\chapter{Conclusions}

In this paper we considered the BJL definition for the
commutators and applied it in conjunction with the operator product
expansion. The method can be applied both in the perturbative and
non-perturbative regimes. As an application of the first case we
considered the anomalous commutator between chiral currents in 1+1 and
3+1 dimensions. We showed that in this case the results in the
literature can be re-interpreted to yield a gauge invariant expression
for the commutators. The method here proposed is consistent with these
results.

In the non-perturbative regime we considered the current algebra
relations between the instanton number density $ \nu $ and the
gauge-invariant anomalous axial charge. We showed that, in general, this
commutator is non-vanishing, in accordance with the results obtained
using instanton calculations. We also noted that this conclusion
is based on the no-trivial chiral transformation properties of the
instanton density and this leads to some modifications of the
expressions resulting from the anomalous Ward identities.

The method requires some knowledge about behaviour of the coefficient
functions (which appear in the OPE) at large momentum transfers $p$. In
asymptotically free theories this is available via the renormalization
group. The final results are expressed in terms of the residues of the
coefficient functions (\ie, the constant multiplying the term behaving
as $ 1/ p_0 $) and of the
matrix elements of various local operators (the ``condensates''). These
quantities can be evaluated explicitly within perturbation theory; in
the non-perturbative regime the condensates cannot be evaluated
explicitly but can be used to parametrize the results.

Whereas the OPE coefficients can be evaluated perturbatively to any
desired order of accuracy, the condensates are not calculable in this
manner; for these quantities effective models must be considered.
Unfortunately the effective theories are valid only at small momenta
and this implies that the $ p^0 \rightarrow \infty $ limit of the OPE
coefficient functions cannot be accurately evaluated using these
theories.
Explicit calculations verify several claims made on general grounds: there
are non-trivial non-canonical contributions to the commutators. These
contributions can be used to reconcile the operator and instanton
approaches to the $\ui_A $ problem.

\ack

J.W. would like to thank R. Akhoury and R. Jackiw for useful comments.
This work was
supported in part through funds provided by the Department of Energy
under contract DE-FG03-94ER40837.

\appendix

\def\ft{\tilde F}

In this appendix we consider the more complicated case of the OPE of the
two current correlator in 3+1 dimensions. In order to keep the
discussion at a manageable level we will consider the case
$$ \tcal^{ab}_{\mu\nu} (p) = \int d^4 x \; e^{ - i p\cdot x }
 T^* J^a_{\mu 5}(x/2) J^b_\nu (-x/2) . \eqn\eq $$
Now $\tcal$ has mass dimension = 2, leading us to
a more complicated OPE. The relevant terms
are~\foot{We use the following conventions:
$\ft^c_{\mu\nu}= \half \epsilon _{\mu\nu\rho\sigma}F_c^{\rho\sigma}$,
with $F^c_{\rho\sigma}
= \partial_\rho A^c_\sigma - \partial_\sigma A^c_\rho + f_{a bc} A^a_\rho
A^b_\sigma$. The covariant derivative is given by
$ \dcal_\mu \ft^c_{\rho\sigma}= \partial_\mu \ft^c_{\rho\sigma} + f_{dec}
A^d_\mu \ft^e_{\rho\sigma}$.}
$$ \eqalign{
\tcal^{ab}_{\mu\nu} (p) = & \inv{p^2}
\Biggl[
u^{a b c }_1 p^\alpha p_\nu \ft^c_{\mu \alpha} +
u^{a b c }_2 p^\alpha p_\mu \ft^c_{\nu \alpha} \cr & \quad +
\left( u^{a b c }_3 \dcal_\mu \ft^c_{ \nu \alpha } +
       u^{a b c }_4 \dcal_\nu \ft^c_{ \mu \alpha } +
       u^{a b c }_5 \dcal_\alpha \ft^c_{ \mu \nu } \right) p^\alpha
\cr & \quad +
u^{a b c }_6 J^c_{ 5 \mu } p_\nu +
u^{a b c }_7 J^c_{ 5 \nu } p_\mu +
u^{a b c }_8 J^{c \rho} p^\sigma \epsilon_{\mu\nu\rho\sigma}
\cr &
\quad + \delta_{ a b } \left(
v_1 J_{ 5 \mu } p_\nu +
v_2 J_{ 5 \nu } p_\mu +
v_3 J^\rho p^\sigma \epsilon_{\mu\nu\rho\sigma} \right) \Biggr] , \cr }
\eqn\eq $$
where $$ u_i^{ a b c } = u_i \up f f_{a b c } + u_i \up d d_{a b c }
\eqn\eq $$
(note that $ \dcal^\alpha \ft_{ \alpha \mu } = 0 $ due to the Bianchi
identities).

Applying the BJL limit to the previous expression, we obtain
$$ \eqalign{
\lim_{p_0 \rightarrow \infty} p_0 \tcal^{ab}_{\mu\nu} & (p) =
i \int d^3 x e^{ i \pp \cdot\xx } \left[ J^a_{\mu 5} (0, \xx/2 ),
J^b_\nu (0 , -\xx/2 ) \right] \cr
= & \left[ u_1^{ a b c } \left( g_{ \nu i } \ft^c_{ \mu 0 } +
                                g_{ \nu 0 } \ft^c_{ \mu i } \right)
         + u_2^{ a b c } \left( g_{ \mu i } \ft^c_{ \nu 0 } +
                                g_{ \mu 0 } \ft^c_{ \nu i } \right) \right] p^i
\cr
& + \left( u^{a b c }_3 \dcal_\mu \ft^c_{ \nu 0 } +
           u^{a b c }_4 \dcal_\nu \ft^c_{ \mu 0 } +
           u^{a b c }_5 \dcal_0 \ft^c_{ \mu \nu } \right) \cr
& + u^{a b c }_6 J^c_{ 5 \mu } g_{ \nu 0 } +
    u^{a b c }_7 J^c_{ 5 \nu } g_{ \mu 0 } +
    u^{a b c }_8 J^{c \rho} \epsilon_{\mu\nu\rho 0 } \cr
& + \delta_{ a b } \left[
v_1 J_{ 5 \mu } g_{ \nu 0 } +
v_2 J_{ 5 \nu } g_{ \mu 0 } +
v_3 J^\rho \epsilon_{\mu\nu\rho 0 } \right] . \cr } \eqn\eq $$

The equal time commutator for the $\mu=\nu=0$ case is given by

$$ \eqalign{
i \left[ J^a_{0 5} (0, \xx/2 ),
J^b_0 (0 , -\xx/2 ) \right] & = \cr
&\mskip-160mu =  - i \left( u_1^{ a b c } + u_2^{ a b c } \right)
          \BB^c \cdot \nabla \delta ( \xx ) \cr
& \mskip-130mu
+ \left[ \left( u^{a b c }_6 + u^{a b c }_7 \right) J^c_{ 5 0 }
+ \delta_{ a b } \left( v_1 + v_2 \right) J_{ 5 0 } \right]  \delta ( \xx ) ,
\cr } \eqn\eq $$
where $ \BB^c $ denotes the chromo-magnetic field, $ B^i = \ft_{ 0 i } $.
The commutator for the space component of the vector current and the time
component of the axial current is
$$ \eqalign{ i \left[ J^a_{0 5} (0, \xx/2 ),
\JJ^b  (0 , -\xx/2 ) \right] = & \cr
& \mskip-160mu = i u_2^{ a b c } \EE^c \times \nabla \delta( \xx ) \cr
& \mskip-140mu + \left[ \left( u^{a b c }_3 - u^{a b c }_5 \right) \dcal_0
\BB^c
+ u^{a b c }_7 \JJ^c_5 + \delta_{ a b } v_2 \JJ_5 \right]
\delta( \xx ) , \cr } \eqn\eq $$
where $ \EE^c $ denotes the chromo-electric field, $ E^i = F_{ 0 i } $.

These results are, as in the 1+1 dimensional case, manifestly gauge
covariant. We have verified that they are consistent with the explicit
loop calculations presented in~\refmark\jo; for example the
two-gauge-boson matrix elements of $ i d_{ c b e } f_{ e a d } \tilde
F_d^{ 0 k } A_k^c $ and $ 8 \pi^2 f_{ a b c } J^c_{ R \mu = 0 } $
coincide.

These results can be used to calculate the corresponding Schwinger terms
and covariantizing
seagulls. Following the procedure described in Ref. \jackiwgross\ we
obtain $ [ J^a_{ \mu \; 5 } ( x/2 )  , J^b_\nu ] \delta ( n \cdot x )
= C_{ \mu \nu }^{ a b }  \delta\up4 ( x ) + S_{ \mu \nu ; \alpha }^{ a b
} \partial^\alpha \delta\up4 ( x ) $ where the Schwinger term
equals (the computation is straightforward and only the results will
be presented) $$ S_{\mu \nu ; \alpha }^{ a b } =
- \left[
u_1^{ a b c } \left(
g_{ \nu \alpha } \tilde F^c_{ \mu \beta } +
g_{ \nu \beta } \tilde F^c_{ \mu \alpha }
\right)
+ u_2^{ a b c } \left(
g_{ \mu \alpha } \tilde F^c_{ \nu \beta } +
g_{ \mu \beta } \tilde F^c_{ \nu \alpha }
\right)
\right] n^\beta \eqn\eq $$ and the corresponding seagull is $$ \tau^{ a b
}_{ \mu \nu } = - \left( u_1^{ a b c } n_\nu \tilde F^c_{ \mu \alpha } +
u_2^{ a b c } n_\mu \tilde F^c_{ \nu \alpha } \right) n^\alpha \eqn\eq $$
We also remark that (again following the procedure described in Ref.
\jackiwgross) when both currents are conserved the requirement
that the $T^*$ product be Lorentz covariant implies $ u_3^{ a b c } =
u_5^{ a b c } $, $ u_1^{ a b c } + u_4^{ a b c } + u_5^{ a b c } = 0 $ and
$ u_6^{ a b c} = u_7^{ a b c} = v_1 = v_2 = 0 $. Finally we note that
the conditions under which the Schwinger terms cancel against the
seagull contributions to the Ward identities is simply $ u_1^{ a b c } +
u_2^{ a b c } = 0 $.

\baselineskip 23 pt
\refout

\bye